\documentclass[12pt]{article}%
\usepackage{graphicx}
\usepackage{amsmath}
\usepackage{amsfonts}
\usepackage{amssymb}%
\providecommand{\U}[1]{\protect\rule{.1in}{.1in}}
\begin{document}

\title{A universal definition of the Kondo energy from the orthogonality catastrophe}
\author{Gerd Bergmann\\Department of Physics\\University of Southern California\\Los Angeles, California 90089-0484\\e-mail: bergmann@usc.edu}
\date{\today}
\maketitle

\begin{abstract}
The definitions of the Kondo energy in the numerical renormalization group
(NRG) and the Friedel artificially inserted resonance (FAIR) theory fail sadly
for small samples where their predicted Kondo energy increases, while in
reality the Kondo effect disappears. Therefore a different, universal
definition of the Kondo energy is proposed, which uses the evasion of the
orthogonality catastrophe by the Kondo impurity. A magnetic impurity which has
a pure diagonal interaction $2Js_{z}S_{z}$ with the conduction electrons
polarizes all of the spin-up and down electrons and reduces the scalar product
between corresponding spin-up and down states. The multi-electron scalar
product (MESP) between all occupied spin-up and spin-down states approaches
zero exponentially with the number $N$ of Wilson states (this is the so-called
orthogonality catastrophe). In contrast in the Kondo ground state the
corresponding conduction electrons of opposite spin are pairwise aligned
within the Kondo energy. In the present paper the MESP is investigated for the
FAIR solution of the Friedel-Anderson impurity. The MESP is numerically
determined for the (enforced) magnetic and the singlet states as a function of
the number $N$ of Wilson states. The magnetic states show an exponentially
decreasing MESP as a function of $N$. Surprisingly it is not the number of
states which causes this decrease. It is instead the smallest energy
separation from the Fermi energy that determines the reduction of the MESP. In
the singlet state the ground state requires a finite MESP to optimize its
energy. As a consequence there is no orthogonality catastrophe. The MESP
approaches a saturation value as function of $N$. Within the energy range of
the Kondo energy the scalar product between corresponding (single electron)
spin-up and spin-down states is very close to $1.000$ and falls off beyond the
Kondo energy. The energy which separates the two regions is well suited as a
universal definition of the Kondo energy.

PACS: 75.20.Hr, 71.23.An, 71.27.+a \newpage

\end{abstract}

\section{Introduction}

There are many definitions of the Kondo temperature in the theoretical solid
state literature. The first one was derived from the divergence of the
perturbation calculation. It yielded essentially
\[
k_{B}T_{K}\thickapprox D\exp\left[  -\frac{1}{2\rho_{0}J}\right]
\]
where $\rho_{0}$ is the density of states, $J$ is the exchange constant in the
interaction $2J\mathbf{s\cdot S,}$ and $D$ is half the band width. The exact
definition depends on the extent to which different families of diagrams are
included in the calculation.

In numerical calculations one generally uses a second generation definition
for the Kondo energy. For example the numerical renormalization group
(\textbf{NRG}) theory defines the Kondo energy $\varepsilon_{K}$ as a quarter
of the inverse susceptibility $\chi\left(  T=0\right)  $ of the spin $1/2$
impurity at zero temperature.
\[
\varepsilon_{K}=\frac{1}{4\chi\left(  T=0\right)  }%
\]
In the FAIR theory (Friedel artificially inserted resonance) our group uses
the singlet-triplet excitation energy. There are two such energies, $E_{st}$
and $E_{st}^{\ast}$. While the first one is just the excitation energy using
the optimal bases in the singlet state the energy $E_{st}^{\ast}$ is obtained
by optimizing the FAIR solution also in the triplet state and using the lowest
possible triplet energy as the excited state. Generally $E_{st}^{\ast}$ is
smaller than $E_{st}$ by a factor not very different from four. On the other
hand the energy $E_{st}$ is almost equal to $\chi^{-1}$. Therefore the NRG and
the FAIR definitions are very close. (Here a 30\% deviation is considered as close).

I consider these definitions of the Kondo energy as second generation because
they cannot be universally applied. One important example where they fail is
small samples. It is well known that the Kondo effect disappears when the
sample size becomes too small (although there were many controversies about
the critical size to show the Kondo effect). The Kondo effect requires a
sufficient number of states in the energy regime $-\varepsilon_{K}%
<\varepsilon<\varepsilon_{K}$ where the energy $\varepsilon$ is measured from
the Fermi level. Although the Kondo effect disappears with decreasing sample
size, neither the susceptibility nor the singlet-triplet excitation energy
disappear. On the contrary both increase with decreasing sample size. They are
obviously not a first class measure for the Kondo energy. Actually our group
ran into this problem while investigating small samples. Therefore we had to
find a more appropriate definition of the Kondo energy. I think we have found
a unique and universal criterion in the way the Kondo effect circumvents the
orthogonality catastrophe. This paper is organized as follows: In chapter II
the orthogonality catastrophe of the Friedel-Anderson (\textbf{FA}) impurity
in the magnetic state is discussed within the frame work of the FAIR theory.
In chapter III numerical results for the multi-electron scalar product are
presented in the enforced magnetic and the singlet state. Its dependence on
the number of states, the smallest energies, and other parameters is
investigated. It turns out that in the Kondo ground state the (single)
electron states for spin up and down are aligned close to the Fermi level
within the Kondo energy. In chapter IV this alignment is investigated
quantitatively and a procedure is developed to extract the Kondo energy
$\varepsilon_{K}$ from this alignment.

\section{Orthogonality catastrophe}

The orthogonality (or infrared) catastrophe was introduced and discussed
already 40 years ago \cite{A53}, \cite{H32}, \cite{Y6}, \cite{Y7}. An example
is a magnetic impurity in a metal host which interacts with the conduction
electron in the form $H^{\prime}=2J\left(  \mathbf{r}\right)  \mathbf{s\cdot
S}$. The effect of the z-component 2$J\left(  \mathbf{r}\right)  s_{z}S_{z}$
is the following. Let the spin direction of the impurity point upwards. Then
the wave function of the conduction electrons is pulled towards or pushed away
from the impurity, depending on the electron spin. As a consequence the scalar
product of corresponding s-electron states with opposite spin is slightly less
than 1.

If we denote the resulting (modified) bases for spin up and down as $\left\{
c_{\nu+}^{\dag}\right\}  $ and $\left\{  c_{\nu-}^{\dag}\right\}  $ with $N$
states in each basis ($1\leq\nu<N$),\ and if half the spin-up and down
sub-bands are occupied then the value of the multi-electron scalar product
(\textbf{MESP}) between all occupied s-states with spin up and those with spin
down is defined by the determinant%
\[
M^{\left(  N/2\right)  }=\left\vert
\begin{array}
[c]{ccc}%
\left\langle c_{1,+}^{\dag}|c_{1,-}^{\dag}\right\rangle  &  & \left\langle
c_{1,+}^{\dag}|c_{N/2,-}^{\dag}\right\rangle \\
&  & \\
\left\langle c_{N/2,+}^{\dag}|c_{1,-}^{\dag}\right\rangle  &  & \left\langle
c_{N/2,+}^{\dag}|c_{N/2-}^{\dag}\right\rangle
\end{array}
\right\vert
\]

The common argument is that the multi-electron scalar product between all
occupied s-states with spin up and those with spin down approaches zero when
$N,$ and therefore the number of occupied s-electron states, becomes very
large. Now we add to this system of spin-up impurity plus polarized conduction
electrons the time reversed system where all spin directions are reversed.
Then the matrix element for a transition between the two states by spin-flip
processes of the form $J\left(  \mathbf{r}\right)  \left[  s^{+}S^{-}%
+s^{-}S^{+}\right]  $ vanishes. Therefore the system cannot decrease its
energy by spin-flip processes. However, at small energy states the gain in
spin-flip energy is larger than the gain in spin polarization. Therefore the
non-diagonal part of the $\mathbf{s\cdot S}$ interaction tries to prevent the
orthogonality catastrophe. This can be well traced in the FAIR treatment of
the Kondo impurity.

In the following the Friedel-Anderson (FA) impurity will be discussed where
this process is less obvious. The Hamiltonian for the FA-impurity is given by
\begin{equation}
H_{FA}=%
%TCIMACRO{\tsum _{\sigma}}%
%BeginExpansion
{\textstyle\sum_{\sigma}}
%EndExpansion
\left\{  \sum_{\nu=1}^{N}\varepsilon_{\nu}c_{\nu,\sigma}^{\dag}c_{\nu,\sigma
}+E_{d}d_{\sigma}^{\dag}d_{\sigma}+\sum_{\nu=1}^{N}V_{sd}(\nu)[d_{\sigma
}^{\dag}c_{\nu,\sigma}+c_{\nu,\sigma}^{\dag}d_{\sigma}]\right\}
+Un_{d\uparrow}n_{d\downarrow} \label{H_FA}%
\end{equation}

In the following I assume that the reader is familiar with the FAIR method
which our group developed during the past few years \cite{B151}, \cite{B152},
\cite{B153}. A short review is posted at the ArXiv \cite{_1}.

Krishna-murthy, Wilkins, and Wilson \cite{K58} clarified the role of the local
magnetic moment in the FA-impurity. They performed a numerical renormalization
a la Wilson \cite{W18} for the FA-Hamiltonian. They demonstrated that for
sufficiently large Coulomb repulsion (when $U>>\Gamma=\pi\rho\left\vert
V_{sd}\right\vert ^{2}$) the flow of their Hamiltonian $H_{N}$ passed close to
the fixed point for a local moment. This means that under these conditions the
impurity first assumed a magnetic moment when the temperature is lowered.
After passing the fixed point for the local moment the renormalization flow
(corresponding to a reduction of temperature) approaches the Kondo ground state.

In the following I will discuss the two different solutions of the
FA-Hamiltonian: the magnetic state and the singlet state. (The magnetic state
can be enforced by a small magnetic field). This state will be called the
\textbf{enforced magnetic state}. This avoids the finite temperature
treatment. In FAIR the magnetic solution $\Psi_{MS}$ has the form
\begin{equation}
\Psi_{MS}=\left[  Aa_{0-\downarrow}^{\dag}a_{0+\uparrow}^{\dag}+Bd_{\downarrow
}^{\dag}a_{0+\uparrow}^{\dag}+Ca_{0-\downarrow}^{\dag}d_{\uparrow}^{\dag
}+Dd_{\downarrow}^{\dag}d_{\uparrow}^{\dag}\right]  \prod_{i=1}^{n-1}%
a_{i-\downarrow}^{\dag}\prod_{i=1}^{n-1}a_{i+\uparrow}^{\dag}\Phi
_{0}\label{PsiMS}%
\end{equation}
The states states $a_{0+}^{\dag}$ and $a_{0-}^{\dag}$ are FAIR states. The
coefficients $A,B,C,D$ and the compositions of the FAIR states are optimized
to minimize the energy expectation value of the FA Hamiltonian. Due to the
condition $\left\langle a_{i\tau}^{\dag}\Phi_{0}\left\vert H_{0}\right\vert
a_{j\tau}^{\dag}\Phi_{0}\right\rangle =0$ the FAIR states determine the other
states $a_{i\tau}^{\dag}$ of the basis $\left\{  a_{i\tau}^{\dag}\right\}  $
uniquely (where $i,j>0$, $H_{0}$ is the free electron Hamiltonian and
$\tau=+,-$ ).

The singlet state is a symmetric superposition of a magnetic state and its
time- (or spin-) reversed state. (The FAIR states $a_{0+}^{\dag}$ and
$a_{0-}^{\dag}$ and the coefficients $A,B,C,D$ are independently optimized for
the magnetic and singlet states).

For the FAIR solution the MESP between the occupied spin up and spin down
sub-bands is essentially given by
\begin{equation}
M_{+-}^{\left(  N/2\right)  }=\left\langle
%TCIMACRO{\tprod \limits_{i=0}^{N/2-1}}%
%BeginExpansion
{\textstyle\prod\limits_{i=0}^{N/2-1}}
%EndExpansion
a_{i+}^{\dag}\Phi_{0}|%
%TCIMACRO{\tprod \limits_{j=0}^{N/2-1}}%
%BeginExpansion
{\textstyle\prod\limits_{j=0}^{N/2-1}}
%EndExpansion
a_{j-}^{\dag}\Phi_{0}\right\rangle \label{MESP}%
\end{equation}%
\[
\]

\section{Numerical Calculation of the Multi-Electron Scalar Product}

\subsection{ The enforced magnetic state}

For most of the numerical calculations Wilson states are used (see appendix).
In the calculation the following parameters are used: $\left\vert
V_{sd}\right\vert ^{2}=0.1$, $U=1,E_{d}=-0.5$. The magnetic solution is
optimized for different numbers of Wilson states with $N=20,30,40,50,60.$
Table I shows $M_{MS}^{\left(  N/2\right)  }$ of the magnetic solution for the
different sizes $N$ of the bases. The third column gives the scalar product of
the two FAIR states, $\left\langle a_{0+}\Phi_{0}\mathbf{|}a_{0-}\Phi
_{0}\right\rangle _{MS},$ the fourth column the ground-state energy, and the
fifth column gives the magnetic moment. ($\Phi_{0}$ is the vacuum state). As
one can see the scalar product $\left\langle a_{0+}\Phi_{0}\mathbf{|}%
a_{0-}\Phi_{0}\right\rangle _{MS},$ the ground-state energy (in the enforced
magnetic state), and the moment have reached their final values already for
$N=30$. However, the multi-scalar product decreases with increasing $N$.%
\[%
\begin{tabular}
[c]{|l|l|l|l|l|l|}\hline
\textbf{N} & $M_{MS}^{\left(  N/2\right)  }$ & $\left\langle a_{0+}%
\mathbf{|}a_{0-}\right\rangle _{MS}$ & $E_{0,MS}$ & $\mu$ & $\left\langle
a_{N/2}\mathbf{|}a_{N/2}\right\rangle _{MS}$\\\hline
10 & 0.878 & 0.823 & -0.607799 & 0.514 & .92\\\hline
20 & 0.396 & 0.501 & -0.627446 & 0.687 & .64\\\hline
30 & 0.190 & 0.4852 & -0.62810 & 0.690 & .55\\\hline
40 & 0.0917 & 0.4845 & -0.62812 & 0.690 & .52\\\hline
50 & 0.0443 & 0.4845 & -0.62812 & 0.690 & .50\\\hline
60 & 0.0216 & 0.484 & -0.62812 & 0.690 & .49\\\hline
2*20 & 0.394 & 0.526 & -0.629323 & 0.66 & .64\\\hline
2*30 & 0.198 & 0.514 & -0.629779 & 0.67 & .57\\\hline
\end{tabular}
\ \ \ \
\]
$%
\begin{tabular}
[c]{l}%
Table I: The multi-electron scalar product (MESP) and other parameters for
the\\
Friedel-Anderson impurity in the enforced magnetic state. The different
columns\\
give the number of Wilson states, the MESP, the (single electron) scalar
product\\
$\left\langle a_{0+}\Phi_{0}\mathbf{|}a_{0-}\Phi_{0}\right\rangle _{MS}$
between the two FAIR states, the ground-state energy, and the\\
magnetic moment. The 6th column is explained in the text. The parameters
used\\
in the calculation are $\left\vert V_{sd}\right\vert ^{2}=0.1$, $U=1,E_{d}%
=-0.5$.
\end{tabular}
\ \ \ \ $%

\[
\]

In Fig.1 the logarithm of the multi-electron scalar product $\ln\left(
M_{MS}^{\left(  N/2\right)  }\right)  $is plotted versus the number of Wilson
states $N.$ It follows a straight line which corresponds to the relation%
\[
M_{MS}^{\left(  N/2\right)  }=1.\,\allowbreak7e^{-0.073\ast N}=1.\,\allowbreak
7\ast0.93^{-N}%
\]
Obviously, the multi-scalar product decreases exponentially with increasing
$N$.%

\[%
\begin{array}
[c]{cc}%
%TCIMACRO{\FRAME{itbpF}{3.1017in}{2.455in}{0in}{}{}{org180_{1}a.eps}%
%{\special{ language "Scientific Word";  type "GRAPHIC";
%maintain-aspect-ratio TRUE;  display "USEDEF";  valid_file "F";
%width 3.1017in;  height 2.455in;  depth 0in;  original-width 3.77in;
%original-height 2.978in;  cropleft "0";  croptop "1";  cropright "1";
%cropbottom "0";  filename 'Org180_1a.eps';file-properties "XNPEU";}}}%
%BeginExpansion
{\includegraphics[
height=2.455in,
width=3.1017in
]%
{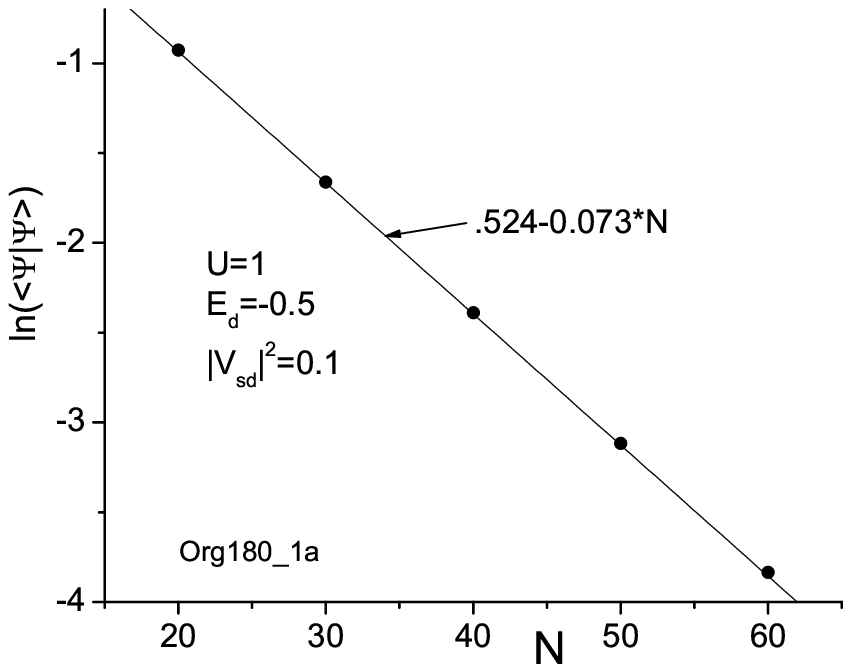}%
}%
%EndExpansion
&
\end{array}
\]

$\
\begin{tabular}
[c]{l}%
Fig.1: The logarithm of the multi-electron scalar product MESP\\
$\left\langle \prod_{i=0}^{n-1}a_{i+}^{\dag}\Phi_{0}|\prod_{i=0}^{n-1}%
a_{i-}^{\dag}\Phi_{0}\right\rangle $ is plotted versus the number of Wilson\\
states $N$ with $n=N/2$ for the magnetic state.
\end{tabular}
\ $%
\[
\]

In the next step I check whether it is just the number of states $N$ which
reduces $M_{MS}^{\left(  N/2\right)  }$. For this purpose the $N$ energy cells
for $N=20$ and $30$ are sub-divided into two. This is achieved by using
$\Lambda=\sqrt{2}.$ This doubles the number of Wilson states but adds only one
state (for positive and negative energy) closer to the Fermi level. The
results of this calculation are collected at $N=2\ast20$ and $2\ast30$. It
turns out that the doubling has essentially only a minor effect on
$M_{MS}^{\left(  N/2\right)  }$. This is on a first sight rather surprising
since it was believed that the increase of the number of states causes the
orthogonality catastrophe of the MESP.

To further confirm this observation I take the energy frame with $N=20$ and
subdivide the energy range $\left(  -1:-1/4\right)  $ into cells with a width
of 1/8, replacing two Wilson states by six new states. (The same is done for
the positive range). This changes $M_{MS}^{N/2}$ from $0.396$ to $0.405$.
Splitting the same energy range into 14 cells with a width of $1/32$ yields
the MESP $M_{MS}^{N/2}=0.411$. This shows that increasing $N$ by subdividing
an energy range does not contribute to an orthogonality catastrophe (as long
as the energy range does not border the Fermi level at the energy $0$).

On the other hand, the smallest (absolute) energies have a great impact on the
MESP. To investigate this question further I take the energies for $N=20$ and
shift the two states which are closest to the Fermi level towards the latter.
The four energy cells which are closest to the Fermi level are $\mathfrak{C}%
_{9}=\left(  -2^{-8}:-2^{-9}\right)  $, $\mathfrak{C}_{10}=\left(
-2^{-9}:0\right)  $, $\mathfrak{C}_{11}=\left(  0:2^{-9}\right)  $,
$\mathfrak{C}_{12}=\left(  2^{-9}:2^{-8}\right)  $. I replace $\pm2^{-9}$ by
$\pm2^{-19}$. Then the (average) energies of the corresponding states are
$\varepsilon_{9}=-\frac{2049}{1048\,576}\thickapprox$ $-1.\,\allowbreak
954\,1\times10^{-3},$ $\varepsilon_{10}=-2^{-20},$ $\varepsilon_{11}=2^{-20}$
and $\varepsilon_{12}=1.\,\allowbreak954\,1\times10^{-3}$. Of course this
reduces the s-d interaction strength $V_{sd}\left(  \nu\right)  $ for
$\nu=10,11$ from $\left[  2^{-9}/2\right]  ^{1/2}=2^{-5}$ to $\left[
2^{-19}/2\right]  ^{1/2}=2^{-10}$.

After optimizing the $\left\{  a_{i+}\right\}  $ and $\left\{  a_{i-}\right\}
$ bases and the coefficients $A,B,C,D$ the resulting MESP is reduced to
$M_{MS}^{10}=0.0208$. The number of states is still $N=20$. The shifting of
the smallest energies from $\pm2^{-10}$ to $\pm2^{-20}$ changes the value of
the MESP from 0.396 to 0.0208. This shows that the value of the MESP is
determined by the occupied state closest to the Fermi level. The total number
of states is only important when it determines the energy of this state.%

\[
\]

\subsection{The singlet state}

In the next step I calculate the MESP for the singlet ground state. The same
parameters $\left\vert V_{sd}\right\vert ^{2}=0.1$, $U=1,E_{d}=-0.5$ are used
as in table I and Fig.1. The FAIR solution for the singlet state is obtained
by reversing all spins in $\Psi_{MS}$ and combining the two states.%
\[
\Psi_{SS}=\Psi_{MS}\left(  \uparrow\downarrow\right)  +\Psi_{MS}\left(
\downarrow\uparrow\right)
\]%
\begin{align}
&  =\left[  Aa_{0-\downarrow}^{\dag}a_{0+\uparrow}^{\dag}+Bd_{\downarrow
}^{\dag}a_{0+\uparrow}^{\dag}+Ca_{0-\downarrow}^{\dag}d_{\uparrow}^{\dag
}+Dd_{\downarrow}^{\dag}d_{\uparrow}^{\dag}\right]  \prod_{i=1}^{n-1}%
a_{i-\downarrow}^{\dag}\prod_{i=1}^{n-1}a_{i+\uparrow}^{\dag}\Phi
_{0}\label{PsiSS}\\
&  +\left[  A^{\prime}a_{0-\uparrow}^{\dag}a_{0+\downarrow}^{\dag}+B^{\prime
}d_{\uparrow}^{\dag}a_{0+\downarrow}^{\dag}+C^{\prime}a_{0-\uparrow}^{\dag
}d_{\downarrow}^{\dag}+D^{\prime}d_{\uparrow}^{\dag}d_{\downarrow}^{\dag
}\right]  \prod_{i=1}^{n-1}a_{i-\uparrow}^{\dag}\prod_{i=1}^{n-1}%
a_{i+\downarrow}^{\dag}\Phi_{0}\nonumber
\end{align}

The coefficients $A,B,C,D,A^{\prime},B^{\prime},C^{\prime},D^{\prime}$ and the
compositions of the FAIR states $a_{0+}^{\dag}$ and $a_{0-}^{\dag}$ are again
optimized to minimize the energy expectation value of the FA Hamiltonian. In
table II the corresponding data are collected. Again the first four columns
give the same data as in table I, i.e. the number of Wilson states, the MESP,
the (single electron) scalar product between the two FAIR states $a_{0+}%
^{\dag}$ and $a_{0-}^{\dag}$ and the ground-state energy. The 5th column gives
the Kondo energy defined as the difference between the relaxed triplet energy
and the singlet ground-state energy. The relaxed  triplet energy is obtained
by setting the primed coefficients $X^{\prime}$ in equ. (\ref{PsiSS}) opposite
equal to the coefficient without prime $X$ and minimizing the energy.

\textbf{Dependence on the number of states }$\mathbf{N}$%

\[%
\begin{tabular}
[c]{|l|l|l|l|l|l|}\hline
\textbf{N} & $M_{SS}^{\left(  N/2\right)  }$ & $\left\langle a_{0+}%
\mathbf{|}a_{0-}\right\rangle _{SS}$ & $E_{0,SS}$ & $\Delta E$ &
%TCIMACRO{\TEXTsymbol{>}}%
%BeginExpansion
$>$%
%EndExpansion
.999\\\hline
10 & 0.749 & 0.645 & -0.62272 & $1\allowbreak4.8\times10^{-3}$ & \\\hline
20 & 0.742 & 0.6448 & -0.637535 & $1\allowbreak0.1\times10^{-3}$ &
9-10\\\hline
30 & 0.742 & 0.6448 & -0.637965 & $9.\,\allowbreak87\times10^{-3}$ &
9-21\\\hline
40 & 0.742 & 0.6448 & -0.63798 & $9.\,\allowbreak86\times10^{-3}$ &
9-31\\\hline
50 & 0.742 & 0.6448 & -0.63798 & $9.\,\allowbreak86\times10^{-3}$ &
9-41\\\hline
60 & 0.742 & 0.6448 & -0.637973 & $9.\,\allowbreak85\times10^{-3}$ &
9-51\\\hline
2*20 & 0.751 & 0.657 & -0.639684 & $1\allowbreak0.4\times10^{-3}$ &
17-23\\\hline
2*30 & 0.751 & 0.657 & -0.639993 & $1\allowbreak0.2\times10^{-3}$ &
18-43\\\hline
\end{tabular}
\]

$%
\begin{tabular}
[c]{l}%
Table II: The multi-electron scalar product (MESP) and other parameters for\\
the Friedel-Anderson impurity in the singlet state. The different columns
give\\
the number of Wilson states $N$, the MESP, the (single electron) scalar
product\\
$\left\langle a_{0+}\Phi_{0}\mathbf{|}a_{0-}\Phi_{0}\right\rangle _{MS}$
between the two FAIR states, the ground-state energy, and the\\
Kondo energy. The 6th column is explained in the text. The parameters used\\
in the calculation are $\left\vert V_{sd}\right\vert ^{2}=0.1$, $U=1,E_{d}%
=-0.5$.
\end{tabular}
\ \ \ $%
\[
\]

For the last column I calculated the scalar product $\left\langle
a_{+,i}^{\dag}\Phi_{0}|a_{-,j}^{\dag}\Phi_{0}\right\rangle $ for all pairs of
$\left(  i,j\right)  $ which form a $N\times N$-matrix. It turns out that the
diagonal elements $\left\langle a_{+,i}^{\dag}\Phi_{0}|a_{-,i}^{\dag}\Phi
_{0}\right\rangle $ close to the Fermi energy approach the value one. For
example\ if the sixth column shows for $N=30$ the value $"9-21"$ then the
values of the (single particle) scalar products $\left\langle a_{+,i}^{\dag
}\Phi_{0}|a_{-,i}^{\dag}\Phi_{0}\right\rangle $ lie between 0.999 and 1.000
for $9\leq i$ $\leq21$. Obviously the states $a_{+,i}^{\dag}$ and
$a_{-,i}^{\dag}$ are almost identical in this interval. This is very different
for the enforced magnetic state. There, in table I the 6th column shows the
value of the diagonal scalar product for $i=N/2$ and the larger one of its two
neighbors $\left\langle a_{+,N/2}^{\dag}\Phi_{0}|a_{-,N/2\pm1}^{\dag}\Phi
_{0}\right\rangle .$

\textbf{Dependence on the interaction }$\left\vert V_{sd}\right\vert ^{2}$

The MESP in the singlet state depends on the strength of the s-d interaction.
Keeping the number of Wilson states constant $N=40,$ the MESP is numerically
determined and collected in table III.
\[%
\begin{tabular}
[c]{|l|l|l|l|l|l|}\hline
$\left\vert \mathbf{V}_{sd}\right\vert ^{2}$ & $M_{SS}^{\left(  N/2\right)  }$
& $\left\langle a_{0+}^{\dag}\mathbf{|}a_{0-}^{\dag}\right\rangle _{SS}$ &
$E_{0,SS}$ & $\Delta E$ &
%TCIMACRO{\TEXTsymbol{>}}%
%BeginExpansion
$>$%
%EndExpansion
.999\\\hline
0.10 & 0.742 & 0.645 & -0.637977 & $9.\,\allowbreak86\times10^{-3}$ &
-\\\hline
0.09 & 0.708 & 0.604 & -0.621392 & $8.\,\allowbreak13\times10^{-3}$ &
10-30\\\hline
0.08 & 0.665 & 0.553 & -0.605078 & $6.\,\allowbreak14\times10^{-3}$ &
10-30\\\hline
0.07 & 0.607 & 0.489 & -0.589200 & $4.\,\allowbreak09\times10^{-3}$ &
11-29\\\hline
0.06 & 0.527 & 0.406 & -0.573975 & $2.\,\allowbreak22\times10^{-3}$ &
11-29\\\hline
0.05 & 0.407 & 0.299 & -0.559655 & $\allowbreak8.\,\allowbreak21\times10^{-4}$
& 12-28\\\hline
\end{tabular}
\ \ \
\]
$%
\begin{tabular}
[c]{l}%
Table III: The multi-electron scalar product (MESP) and other parameters for\\
the Friedel-Anderson impurity in the singlet state. The columns give the\\
s-d interaction $\left\vert V_{sd}\right\vert ^{2}$, the MESP, the scalar
product $\left\langle a_{0+}^{\dag}\Phi_{0}|a_{0-}^{\dag}\phi_{0}\right\rangle
$, the\\
ground-state energy, and Kondo energy $\Delta E.$ The 6th column is explained
in\\
the text. The parameters used in the calculation are $U=1,E_{d}=-0.5$ and\\
the number of Wilson states is $N=40$.
\end{tabular}
\ \ \ $%
\[
\]

In Fig.2 the logarithm of the Kondo energy is plotted versus the logarithm of
the MESP. A linear dependence is obtained. The MESP shows a weak dependence on
the Kondo energy with a power of about $1/4$.%

\[%
\begin{array}
[c]{cc}%
%TCIMACRO{\FRAME{itbpF}{3.2013in}{2.6359in}{0in}{}{}{org180_{4}a.eps}%
%{\special{ language "Scientific Word";  type "GRAPHIC";
%maintain-aspect-ratio TRUE;  display "USEDEF";  valid_file "F";
%width 3.2013in;  height 2.6359in;  depth 0in;  original-width 3.77in;
%original-height 3.0992in;  cropleft "0";  croptop "1";  cropright "1";
%cropbottom "0";  filename 'Org180_4a.eps';file-properties "XNPEU";}}}%
%BeginExpansion
{\includegraphics[
height=2.6359in,
width=3.2013in
]%
{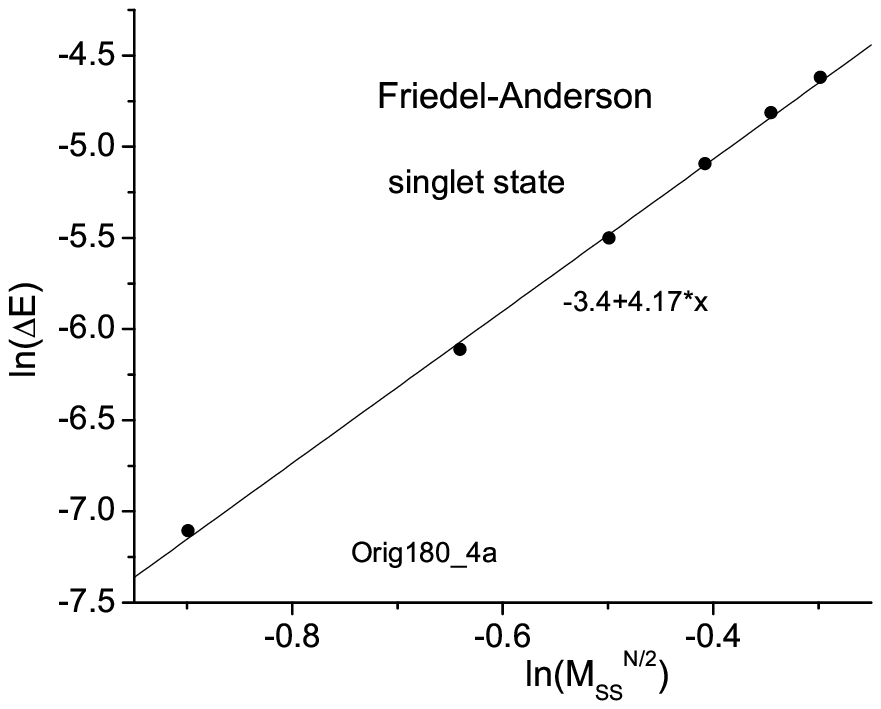}%
}%
%EndExpansion
&
\end{array}%
\begin{tabular}
[c]{l}%
Fig.2: The log-log plot of the\\
Kondo energy versus the MESP\\
for the singlet state for different\\
s-d interactions.
\end{tabular}
\ \
\]
$\ \ \ $%
\[
\]

\section{State alignment within the Kondo energy}

\subsection{Friedel-Anderson impurity}

The states $a_{+,i}^{\dag}$ are constructed from the basis $c_{\nu}^{\dag}$ by
extracting a FAIR state $a_{+,0}^{\dag}$. Therefore the states $a_{+,i}^{\dag
}$ and $c_{\nu}^{\dag}$ are pairwise quite similar except that there is one
state missing in the basis $\left\{  a_{+,i}\right\}  $. The same applies to
the states $a_{-,i}^{\dag}$ and $c_{\nu}^{\dag}.$
\[%
\begin{array}
[c]{cc}%
%TCIMACRO{\FRAME{itbpF}{3.4811in}{2.7223in}{0in}{}{}{sclprdfa_{n}%
%50a.eps}{\special{ language "Scientific Word";  type "GRAPHIC";
%maintain-aspect-ratio TRUE;  display "USEDEF";  valid_file "F";
%width 3.4811in;  height 2.7223in;  depth 0in;  original-width 3.7825in;
%original-height 2.9523in;  cropleft "0";  croptop "1";  cropright "1";
%cropbottom "0";  filename 'SclPrdFA_N50a.eps';file-properties "XNPEU";}}}%
%BeginExpansion
{\includegraphics[
height=2.7223in,
width=3.4811in
]%
{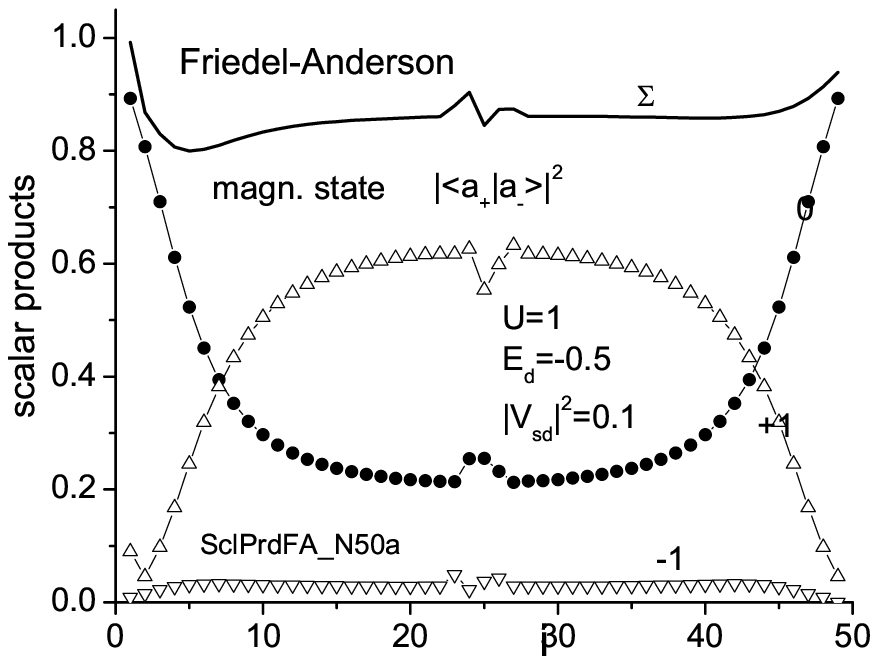}%
}%
%EndExpansion
&
\end{array}%
\begin{tabular}
[c]{l}%
Fig.4: The square of the diagonal\\
matrix-elements $\left\vert \left\langle a_{+,i}^{\dag}\Phi_{0}|a_{-,i}^{\dag
}\Phi_{0}\right\rangle \right\vert ^{2}$\\
as well as the next-to-diagonal\\
matrix-elements $\left\vert \left\langle a_{+,i}^{\dag}\Phi_{0}|a_{-,i\pm
1}^{\dag}\Phi_{0}\right\rangle \right\vert ^{2}$\\
are plotted as a function of $i$ for the\\
enforced magnetic state.
\end{tabular}
\ \ \ \
\]

\[
\]

As a consequence the two bases $\left\{  a_{+,i}^{\dag}\right\}  $ and
$\left\{  a_{-,i}^{\dag}\right\}  $ are also quite similar. In Fig.4 the
single-particle scalar products $\left\vert \left\langle a_{+,i}^{\dag}%
\Phi_{0}|a_{-,i}^{\dag}\Phi_{0}\right\rangle \right\vert ^{2}$ for the
enforced magnetic state are plotted as a function of $i$ (full circles). In
addition $\left\vert \left\langle a_{+,i}^{\dag}\Phi_{0}|a_{-,i\pm1}^{\dag
}\Phi_{0}\right\rangle \right\vert ^{2}$ are plotted as empty up and down
triangles. The full curve (without symbols) gives the sum of the three
contributions. One recognizes that an arbitrary state $a_{+,i}^{\dag}$ (for
$i>0$) consists 80\% out of the states $a_{-,i}^{\dag},a_{-,i+1}^{\dag}$ and
$a_{-,i-1}^{\dag}$. On the other hand $a_{+,i}^{\dag}$ and $a_{-,i}^{\dag}$
overlap only 30\% for small energies (in the center of the horizontal axis).

This is very different for the singlet state. In Fig.5 the single-electron
scalar products $\left\vert \left\langle a_{+,i}^{\dag}\Phi_{0}|a_{-,i}^{\dag
}\Phi_{0}\right\rangle \right\vert ^{2}$ as well as $\left\vert \left\langle
a_{+,i}^{\dag}\Phi_{0}|a_{-,i\pm1}^{\dag}\Phi_{0}\right\rangle \right\vert
^{2}$ are plotted as a function of $i$ for the singlet state. One recognizes
that over a large energy range the states $a_{+,i}^{\dag}$ and $a_{-,i}^{\dag
}$ are 99\% or more identical. Only for (absolute) large energies on the left
and right side is the overlap reduced to about 70\%.%
\[%
\begin{array}
[c]{cc}%
%TCIMACRO{\FRAME{itbpF}{3.7642in}{2.9439in}{0in}{}{}{sclprdfa_{n}%
%50b.eps}{\special{ language "Scientific Word";  type "GRAPHIC";
%maintain-aspect-ratio TRUE;  display "USEDEF";  valid_file "F";
%width 3.7642in;  height 2.9439in;  depth 0in;  original-width 3.7825in;
%original-height 2.9523in;  cropleft "0";  croptop "1";  cropright "1";
%cropbottom "0";  filename 'SclPrdFA_N50b.eps';file-properties "XNPEU";}}}%
%BeginExpansion
{\includegraphics[
height=2.9439in,
width=3.7642in
]%
{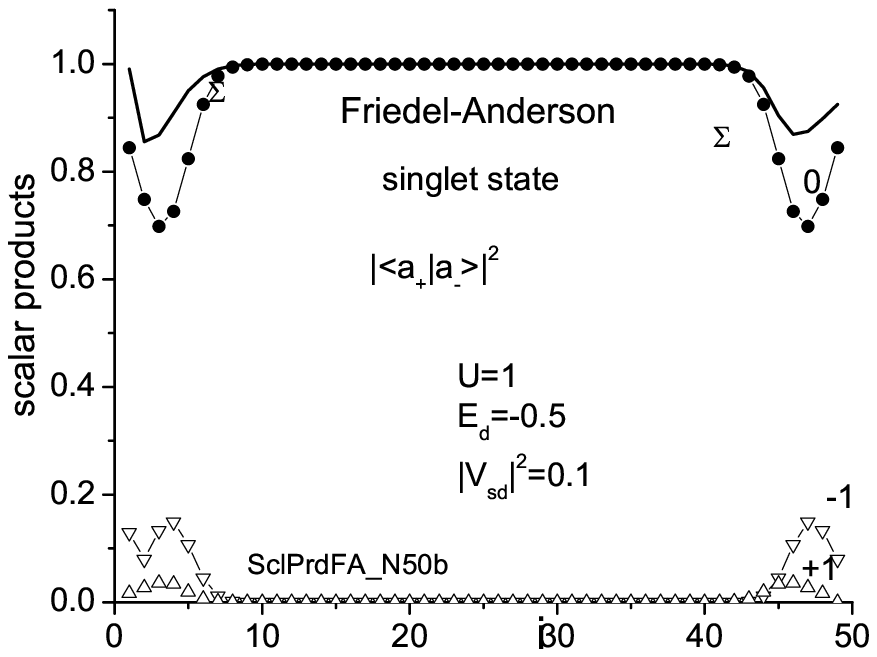}%
}%
%EndExpansion
&
\end{array}%
\begin{tabular}
[c]{l}%
Fig.5: The square of the diagonal\\
matrix-elements $\left\vert \left\langle a_{+,i}^{\dag}\Phi_{0}|a_{-,i}^{\dag
}\Phi_{0}\right\rangle \right\vert ^{2}$\\
as well as the next-to-diagonal\\
matrix-elements $\left\vert \left\langle a_{+,i}^{\dag}\Phi_{0}|a_{-,i\pm
1}^{\dag}\Phi_{0}\right\rangle \right\vert ^{2}$\\
are plotted as a function of $i$ for the\\
singlet state.
\end{tabular}
\ \ \ \ \ \
\]

\[
\]

The reason for this different behavior is rather transparent. The energy
expectation value of the enforced magnetic state does not depend on the MESP.
The s-d transitions happen only within the same spin orientation and therefore
within the same basis. There is no advantage of having the states
$a_{+,i}^{\dag}$ and $a_{-,i}^{\dag}$ aligned.

On the other hand in the singlet state one has transitions from
\[
a_{0-\downarrow}^{\dag}a_{0+\uparrow}^{\dag}\prod_{i=1}^{n-1}a_{i-\downarrow
}^{\dag}\prod_{i=1}^{n-1}a_{i+\uparrow}^{\dag}\Phi_{0}<=>d_{\uparrow}^{\dag
}a_{0+\downarrow}^{\dag}\prod_{i=1}^{n-1}a_{i-\uparrow}^{\dag}\prod
_{i=1}^{n-1}a_{i+\downarrow}^{\dag}\Phi_{0}%
\]
Such a transition is proportional to the square of the MESP. To be able to
harvest energy from these processes the states $a_{i+}^{\dag}$ and
$a_{i-}^{\dag}$ are, for small energies, aligned with each other.

The next question investigated is how the alignment of the states
$a_{i+}^{\dag}$ and $a_{i-}^{\dag}$ depends on the energy $E_{i,\pm}$ of the
states $a_{i+}^{\dag}$ and $a_{i-}^{\dag}$. In Fig.6 the scalar product
$\left\langle a_{i+}^{\dag}\Phi_{0}|a_{i-}^{\dag}\Phi_{0}\right\rangle $ is
plotted as a function of $\log\left\vert E_{i,\pm}\right\vert $ for the
parameters $\left\vert V_{sd}\right\vert ^{2}=0.04,$ $E_{d}=-0.5$ and $U=1.0$.
One recognizes that for positive and negative energies the scalar product at
small energies is essentially equal to one. Then at larger energies the values
of the scalar product decrease in first approximation linearly. The
intersection of the two straight lines is at $\log\left\vert E_{i,\pm
}\right\vert =-3.41$. The corresponding energy of $3.\,\allowbreak
90\times10^{-4}$ can be used as a new definition for the Kondo energy.%

\[%
\begin{array}
[c]{c}%
%TCIMACRO{\FRAME{itbpF}{2.8908in}{2.3968in}{0in}{}{}{org180_{5}a.eps}%
%{\special{ language "Scientific Word";  type "GRAPHIC";
%maintain-aspect-ratio TRUE;  display "USEDEF";  valid_file "F";
%width 2.8908in;  height 2.3968in;  depth 0in;  original-width 3.7443in;
%original-height 3.0992in;  cropleft "0";  croptop "1";  cropright "1";
%cropbottom "0";  filename 'Org180_5a.eps';file-properties "XNPEU";}}}%
%BeginExpansion
{\includegraphics[
height=2.3968in,
width=2.8908in
]%
{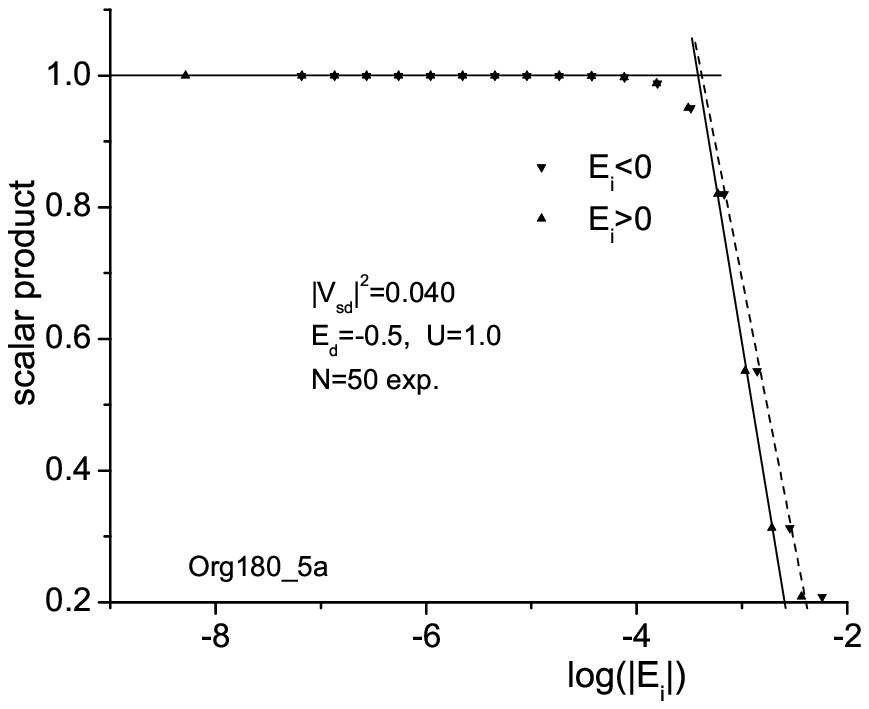}%
}%
%EndExpansion
\end{array}%
\begin{tabular}
[c]{l}%
Fig.6: The scalar product $\left\langle a_{+,i}^{\dag}\Phi_{0}|a_{-,i}^{\dag
}\Phi_{0}\right\rangle $\\
as a function of the logarithm of the\\
absolute value of the energies $\log\left\vert E_{i,\pm}\right\vert $.\\
The parameters are $\left\vert V_{sd}\right\vert ^{2}=0.04,$\\
$E_{d}=-0.5$ and $U=1.0$ for the\\
Friedel-Anderson impurity.
\end{tabular}
\]

In Fig.7 a similar plot is shown for different values of $\left\vert
V_{sd}\right\vert ^{2}$. The smaller the value of $\left\vert V_{sd}%
\right\vert ^{2}$ the smaller is the energy of the intersection. For the value
of $0.02$ the scalar product never reaches the value one. For this value the
Kondo energy is less than $10^{-9}$. For $N=50$ the smallest energy of the
states $c_{\nu}^{\dag}$ (and correspondingly for $a_{i,\pm}^{\dag}$) is
$2^{-25}\thickapprox\allowbreak3\times10^{-8}$. This corresponds also to a
$\Delta k=\left\vert k-k_{F}\right\vert \thickapprox3\times10^{-8}$ ($k$ is
the wave number is the corresponding state). Through the uncertainty principle
(or through the quantization condition $\Delta k=\pi/R$) this corresponds to a
finite sample radius of $R\thickapprox1.\times10^{8}$ (see appendix).%

\[%
\begin{array}
[c]{c}%
%TCIMACRO{\FRAME{itbpF}{3.1557in}{2.5645in}{0in}{}{}{org180_{5}b.eps}%
%{\special{ language "Scientific Word";  type "GRAPHIC";
%maintain-aspect-ratio TRUE;  display "USEDEF";  valid_file "F";
%width 3.1557in;  height 2.5645in;  depth 0in;  original-width 3.8373in;
%original-height 3.1133in;  cropleft "0";  croptop "1";  cropright "1";
%cropbottom "0";  filename 'Org180_5b.eps';file-properties "XNPEU";}}}%
%BeginExpansion
{\includegraphics[
height=2.5645in,
width=3.1557in
]%
{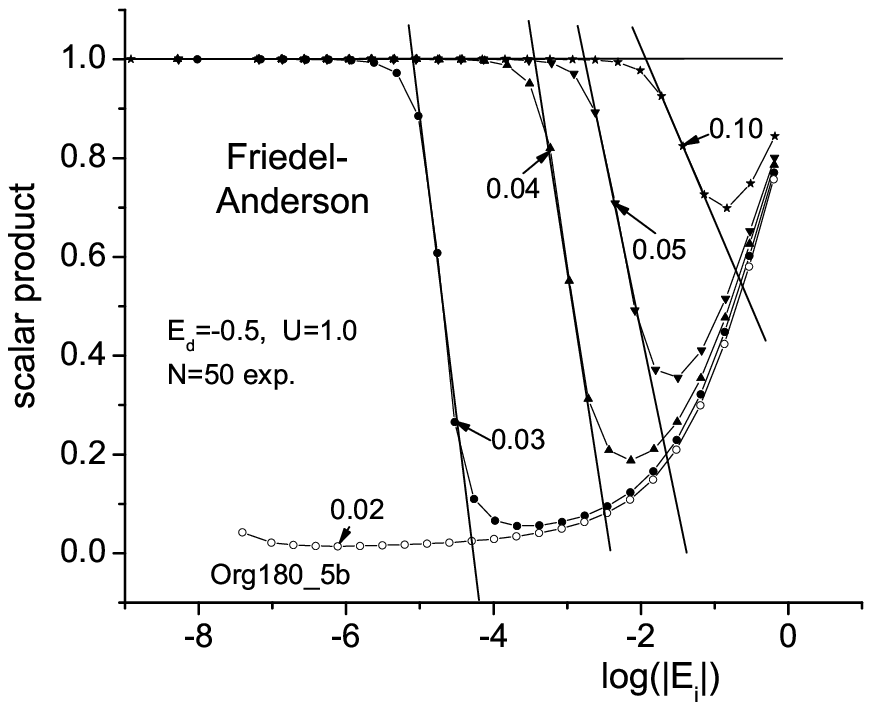}%
}%
%EndExpansion
\end{array}%
\begin{tabular}
[c]{l}%
Fig.7: The scalar product $\left\langle a_{+,i}^{\dag}\Phi_{0}|a_{-,i}^{\dag
}\Phi_{0}\right\rangle $\\
as a function of $\log\left\vert E_{i,+}\right\vert $ for the different\\
s-d-coupling in the Friedel-Anderson\\
impurity. The value of \ $\left\vert V_{sd}\right\vert ^{2}\ \ $is shown\\
next to the curves. The d-state and Coulomb\\
energies are $E_{d}=-0.5$ and $U=1.0$.
\end{tabular}
\
\]

In Fig.8 the different results are collected. Here the extrapolated logarithm
of the energy $\log\left(  \varepsilon_{K}\right)  $ is used as abscissa. For
each parameter set the logarithm of the unrelaxed singlet-triplet energy
$\log\left(  E_{st}\right)  $ (full circle), the relaxed singlet-triplet
energy $\log\left(  E_{st}^{\ast}\right)  $ (full triangle) and the
susceptibility energy $\log\left(  E_{\chi}\right)  $ (stars) are plotted).
Here $E_{\chi}$ is defined by the inverse susceptibility as $E_{\chi
}=1/\left(  4\chi\right)  $. (The latter has been recently calculated for the
FAIR approach \cite{B182}.) Along the straight line abscissa and ordinate are
equal. The different definitions for the Kondo energy yield rather similar
results. The new definition yields slightly larger values than the
susceptibility and the relaxed $E_{st}^{\ast}$, but smaller values than the
unrelaxed one $E_{st}$.
\begin{align*}
&
\begin{array}
[c]{c}%
%TCIMACRO{\FRAME{itbpF}{3.2312in}{2.7314in}{0in}{}{}{org180_{5}c.eps}%
%{\special{ language "Scientific Word";  type "GRAPHIC";
%maintain-aspect-ratio TRUE;  display "USEDEF";  valid_file "F";
%width 3.2312in;  height 2.7314in;  depth 0in;  original-width 3.7825in;
%original-height 3.1922in;  cropleft "0";  croptop "1";  cropright "1";
%cropbottom "0";  filename 'Org180_5c.eps';file-properties "XNPEU";}}}%
%BeginExpansion
{\includegraphics[
height=2.7314in,
width=3.2312in
]%
{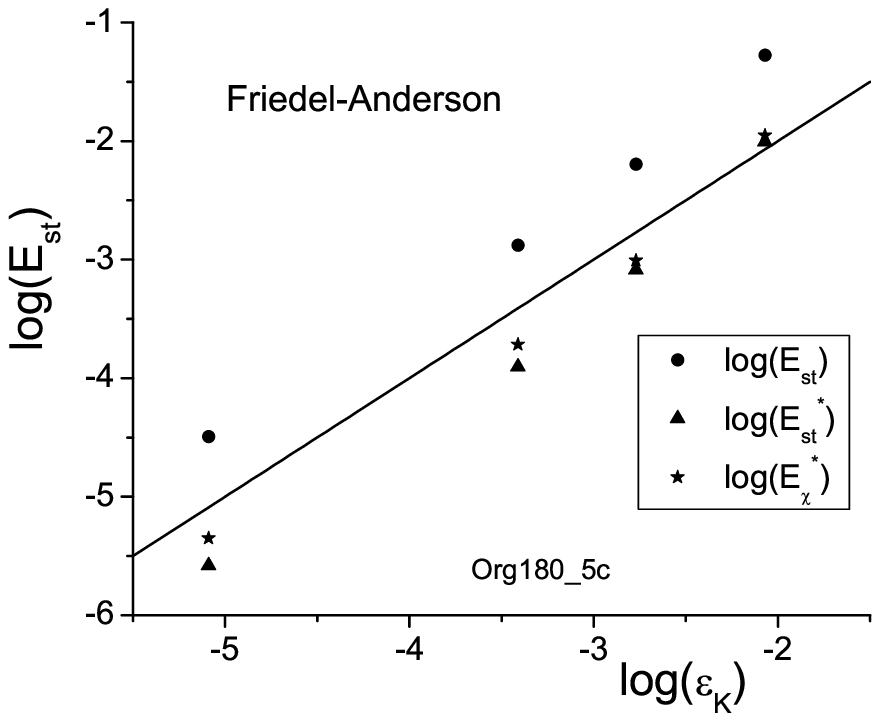}%
}%
%EndExpansion
\end{array}
\\
&
\begin{tabular}
[c]{l}%
Fig.8: The logarithm of the unrelaxed energy $E_{st}$ (full circle),\\
relaxed $E_{st}^{\ast}$ (triangle) and susceptibility energy $E_{\chi}$ as a\\
function of the logarithm of the extrapolated energy $\varepsilon_{K}$. The\\
straight line corresponds to $\log\left(  E_{st}\right)  =\log\left(
\varepsilon_{K}\right)  $.
\end{tabular}
\end{align*}

\subsection{Kondo impurity}

The same calculations are performed for the Kondo impurity. In Fig.9 a similar
plot is shown for different values of $J$. The smaller the value of $J$ the
smaller is the energy of the intersection.
\[%
\begin{array}
[c]{c}%
%TCIMACRO{\FRAME{itbpF}{3.1474in}{2.5579in}{0in}{}{}{org180_{6}b.eps}%
%{\special{ language "Scientific Word";  type "GRAPHIC";
%maintain-aspect-ratio TRUE;  display "USEDEF";  valid_file "F";
%width 3.1474in;  height 2.5579in;  depth 0in;  original-width 3.8373in;
%original-height 3.1133in;  cropleft "0";  croptop "1";  cropright "1";
%cropbottom "0";  filename 'Org180_6b.eps';file-properties "XNPEU";}}}%
%BeginExpansion
{\includegraphics[
height=2.5579in,
width=3.1474in
]%
{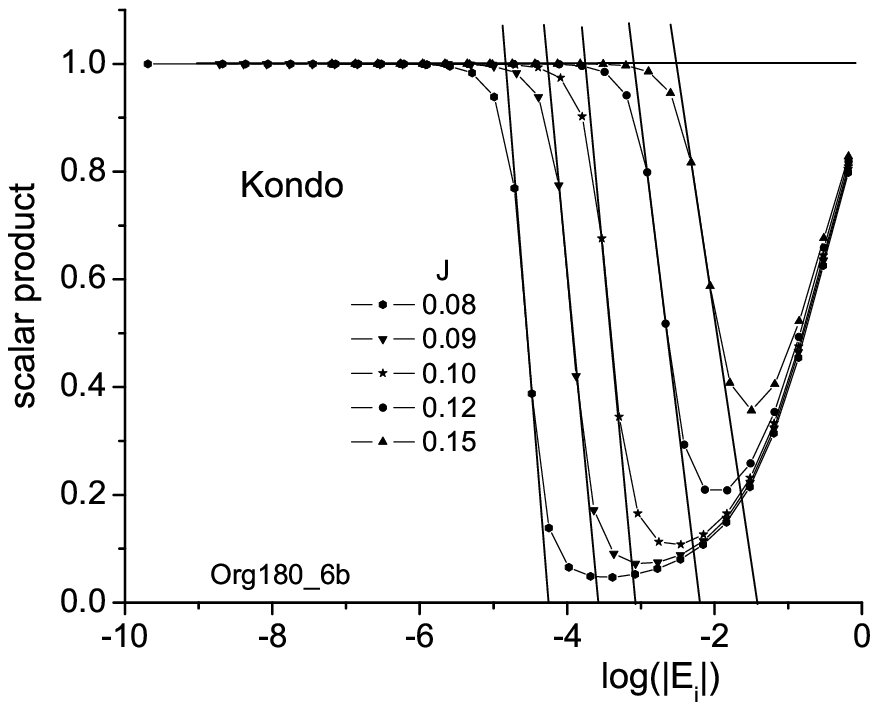}%
}%
%EndExpansion
\end{array}%
\begin{tabular}
[c]{l}%
Fig.9: The scalar product $\left\langle a_{+,i}^{\dag}\Phi_{0}|a_{-,i}^{\dag
}\Phi_{0}\right\rangle $\\
as a function of $\log\left\vert E_{i,+}\right\vert $ for the different\\
s-d-coupling in the Kondo impurity. The\\
values of $J\ $are shown next to the curves.
\end{tabular}
\ \
\]

In Fig.10 the different results are collected. Again the extrapolated
logarithm of the energy $\varepsilon_{K}$ is used as abscissa. For each
parameter set the logarithm of the unrelaxed singlet-triplet energy
$\log\left(  E_{st}\right)  $ (full circle), the relaxed singlet-triplet
energy $\log\left(  E_{st}^{\ast}\right)  $ (full triangle) and the
susceptibility energy $\log\left(  E_{\chi}\right)  $ (stars) are plotted).
Again the different values are essentially proportional to each other.%
\begin{align*}
&
\begin{array}
[c]{c}%
%TCIMACRO{\FRAME{itbpF}{3.3391in}{2.6559in}{0in}{}{}{org180_{6}c.eps}%
%{\special{ language "Scientific Word";  type "GRAPHIC";
%maintain-aspect-ratio TRUE;  display "USEDEF";  valid_file "F";
%width 3.3391in;  height 2.6559in;  depth 0in;  original-width 3.8373in;
%original-height 3.0461in;  cropleft "0";  croptop "1";  cropright "1";
%cropbottom "0";  filename 'Org180_6c.eps';file-properties "XNPEU";}}}%
%BeginExpansion
{\includegraphics[
height=2.6559in,
width=3.3391in
]%
{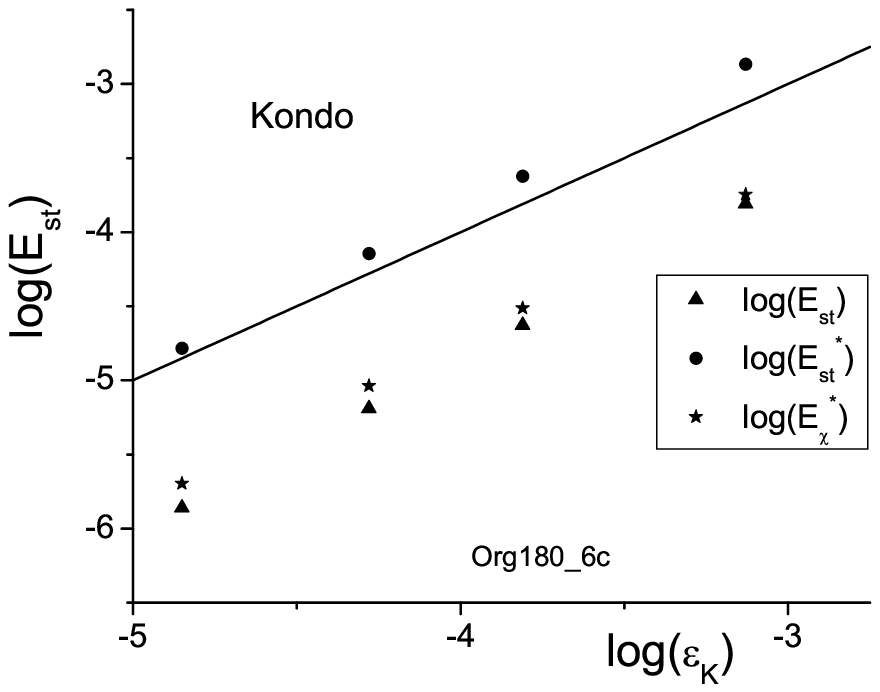}%
}%
%EndExpansion
\end{array}
\\
&
\begin{tabular}
[c]{l}%
Fig.10: The logarithm of the unrelaxed energy $E_{st}$ (full circle),\\
relaxed $E_{st}^{\ast}$ (triangle) and susceptibility energy $E_{\chi}$ as a\\
function of the logarithm of the extrapolated energy $\varepsilon_{K}$.
\end{tabular}
\end{align*}
\[
\]

\section{Conclusion}

The Kondo effect develops only in samples of sufficiently large size. This is
(at least) known since Wilson's NRG paper. However, two of the measures for
the Kondo energy, the suceptibility energy and the singlet-triplet excitation
energy increase when the sample size is decreased. Both fail as a measure for
the Kondo energy for small sample sizes. In the present paper the alignment of
the electronic wave functions close to the Fermi energy is proposed as an
alternative measure for the Kondo energy. This alignment is intimately
connected to the Kondo ground state. In the Kondo ground state the alignment
takes place to avoid the orthogonality catastrophe. The latter prevents any
energy gain of the ground-state energy from spin-flip processes. Therefore the
energy range in which spin-up and down electron states are aligned is also the
energy range in which spin-flip processes contribute to the reduction of the
ground-state energy. Since this energy range is exponentially small the Kondo
energy is exponentially small.

A satisfactory definition of the Kondo energy that shows the disappearance of
the Kondo effect is required to investigate the presence (or absence) of the
Kondo effect in small samples, in particular in three dimensions. It was long
believed that the critical size for the Kondo effect in three dimensions is
the Kondo radius $r_{K}=\hbar v_{F}/\varepsilon_{K}$. (In Wilson nomiclature
$\hbar v_{F}$ is equal to one). The author \cite{B59} investigated the
conditions for the development of a resonance in three dimensions. Only a
perfect sphere with the impurity in the center requires such a large size as
$\hbar v_{F}/\varepsilon_{R}$ (where $\varepsilon_{R}$ is the energy width of
the resonance). For less symmetric samples a much smaller size is sufficient.
Presently our group is investigating this question for the Kondo resonance.
The criterion developed here for the Kondo energy and the occurrence of the
Kondo effect is essential for this investigation.%
\[
\]

\section{Appendix}

\appendix{}

\section{Kondo Effect in Small Samples}

As mentioned above one can simulate a small sample by using a finite number of
Wilson states. Wilson \cite{W18} in his Kondo paper considered an s-band
ranging from $-1$ to 1 with a constant density of states. Then Wilson replaced
the energy continuum of s-states by a discrete set of cells. First the
negative energy band is subdivided on a logarithmic scale. The discrete energy
values are $-1,-1/\Lambda,-1/\Lambda^{2}$,$-\Lambda^{-\nu},$ $..-\Lambda
^{-\left(  N/2-1\right)  },0$. (More often than not the value chosen for
$\Lambda$ is $2$). These discrete points $\xi_{\nu}=-\Lambda^{-\nu}$ are used
to define a sequence of energy cells: the cell $\mathfrak{C}_{\nu}$ (for $\nu$%
%TCIMACRO{\TEXTsymbol{<}}%
%BeginExpansion
$<$%
%EndExpansion
$N/2$) includes all states within $\left(  \xi_{\nu-1}:\xi_{\nu}\right)
=\left(  -1/\Lambda^{\nu-1}:-1/\Lambda^{\nu}\right)  $. A new (Wilson) state
$c_{\nu}^{\dag}$ is a superposition of all states within an energy cell
$\left(  \xi_{\nu-1}:\xi_{\nu}\right)  $ and has an (averaged) energy $\left(
\xi_{\nu-1}+\xi_{\nu}\right)  /2=\allowbreak\left(  -\frac{\Lambda+1}%
{2}\right)  \dfrac{1}{\Lambda^{\nu}}$. This yields for $\Lambda=2$ a spectrum
$\varepsilon_{\nu}$: $-\frac{3}{4},-\frac{3}{8},-\frac{3}{16},$ $..,-\frac
{3}{2^{N/2}},-\frac{1}{2^{N/2}}$. This spectrum is extended symmetrically to
positive energies (for $\nu>N/2$). For a given $N$ the two smallest energy
cells extend from $\pm2^{-\left(  N/2-1\right)  }$ to $0,$ and the (absolute)
smallest energy levels are $\pm2^{-N/2}$. The (absolute) smallest energies
(with respect to the Fermi level) correspond to a smallest wave number $\Delta
k$ which is equal to $\Delta\varepsilon$ because Wilson uses the dispersion
relation $\varepsilon=\left(  k-1\right)  $. Therefore one has $\Delta
k=\left\vert k-k_{F}\right\vert \thickapprox2^{-N/2}$. Since the smallest
$\Delta k$ is connected with the finite size of the sample through the
relation
\[
R=\alpha_{d}\frac{\pi}{\Delta k}%
\]
where $R$ is the radius of the sample and $\alpha_{d}$ is in one dimension
$\alpha_{1}=1/2$ and in three dimensions $\alpha_{3}=1$. Therefore one can
simulate a sample of finite size by a small number $N$ of Wilson states. It is
well known from Wilson's NRG theory that the Kondo effect occurs only for a
sufficiently large $N$, i.e. a sufficiently large sample where the sample size
is larger than the Kondo length $R_{K}=\hbar v_{F}/\varepsilon_{K}$.

In Fig.11 the scalar products $\left\langle a_{+,i}^{\dag}\Phi_{0}%
|a_{-,i}^{\dag}\Phi_{0}\right\rangle $ are plotted as a function of
$\log\left\vert E_{i,+}\right\vert $ for different $N$ or sample size. The
curves for $N=20$ (open circles) and $N=30$ ($\triangle$) don't come close to
the value of one at the smallest value of $E_{i,+}$. For $N=40$
($\triangledown$) the curve just reaches the value of one and for $N=50$
(stars) the curve assumes the value one over two decades of energy.%

\begin{align*}
&
\begin{array}
[c]{c}%
%TCIMACRO{\FRAME{itbpF}{3.9493in}{3.2179in}{0in}{}{}{orig180_{7}a.eps}%
%{\special{ language "Scientific Word";  type "GRAPHIC";
%maintain-aspect-ratio TRUE;  display "USEDEF";  valid_file "F";
%width 3.9493in;  height 3.2179in;  depth 0in;  original-width 3.7443in;
%original-height 3.0461in;  cropleft "0";  croptop "1";  cropright "1";
%cropbottom "0";  filename 'Orig180_7a.eps';file-properties "XNPEU";}}}%
%BeginExpansion
{\includegraphics[
height=3.2179in,
width=3.9493in
]%
{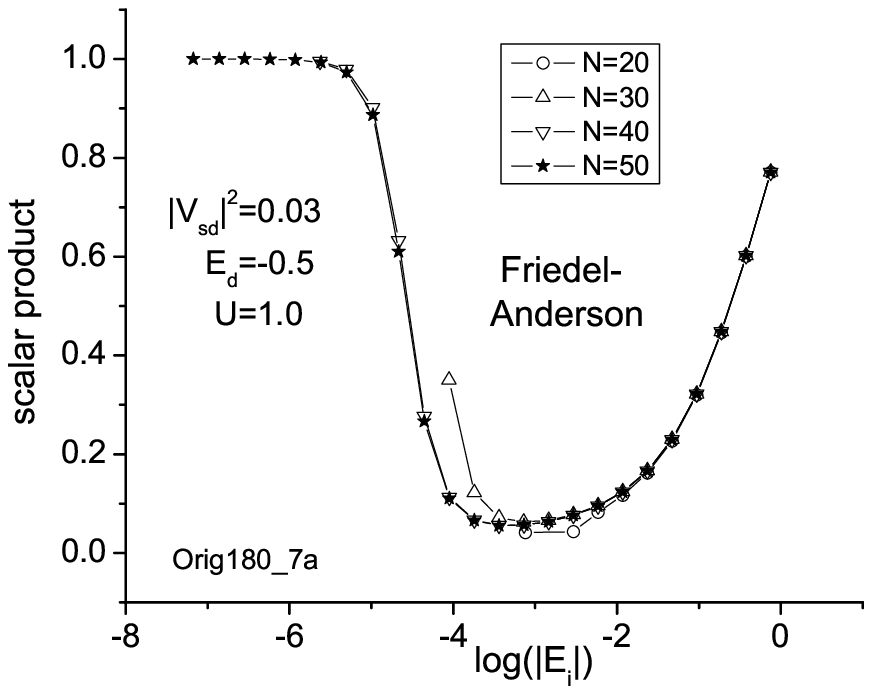}%
}%
%EndExpansion
\end{array}
\\
&
\begin{tabular}
[c]{l}%
Fig.11: The scalar product $\left\langle a_{+,i}^{\dag}\Phi_{0}|a_{-,i}^{\dag
}\Phi_{0}\right\rangle $ as a function of $\log\left\vert E_{i,+}\right\vert
$\\
for the different number $N$ of Wilson states for a Friedel-Anderson
impurity.\\
The parameters are \ $\left\vert V_{sd}\right\vert ^{2}=0.03,\ E_{d}=-0.5$ and
$U=1.0$. Only the\\
curves for $N=40$ and $50$ reach the value close to one.
\end{tabular}
\end{align*}

In Fig.12 the value of the (logarithm of the) singlet-triplet excitation
energy $E_{st}$ and the corresponding Kondo energy from the susceptibility
($E_{\chi}=1/4\chi$ ) are plotted versus the number $N$ of Wilson states. The
Kondo energy is of the order of $10^{-5}\thickapprox2^{-16.5}$. This
corresponds to an $N_{cr}\thickapprox33.$ And indeed one recognizes that the
energies $E_{st}$ and $E_{\chi}$ increase for $N<33\thickapprox N_{cr}$. This
value corresponds to a sample size of $R\thickapprox2^{16.5}\thickapprox
10^{5}.$For smaller samples ($N<N_{cr}$) the two expressions for the Kondo
energy loose their meaning. They don't indicate by themselves that the Kondo
effect has disappeared. On the other hand the spin-up and down states close to
the Fermi energy have lost their alignment (see Fig.11), showing that the
Kondo effect is destroyed.%

\begin{align*}
&
\begin{array}
[c]{c}%
%TCIMACRO{\FRAME{itbpF}{4.005in}{3.2179in}{0in}{}{}{orig180_{7}b.eps}%
%{\special{ language "Scientific Word";  type "GRAPHIC";
%maintain-aspect-ratio TRUE;  display "USEDEF";  valid_file "F";
%width 4.005in;  height 3.2179in;  depth 0in;  original-width 3.7966in;
%original-height 3.0461in;  cropleft "0";  croptop "1";  cropright "1";
%cropbottom "0";  filename 'Orig180_7b.eps';file-properties "XNPEU";}}}%
%BeginExpansion
{\includegraphics[
height=3.2179in,
width=4.005in
]%
{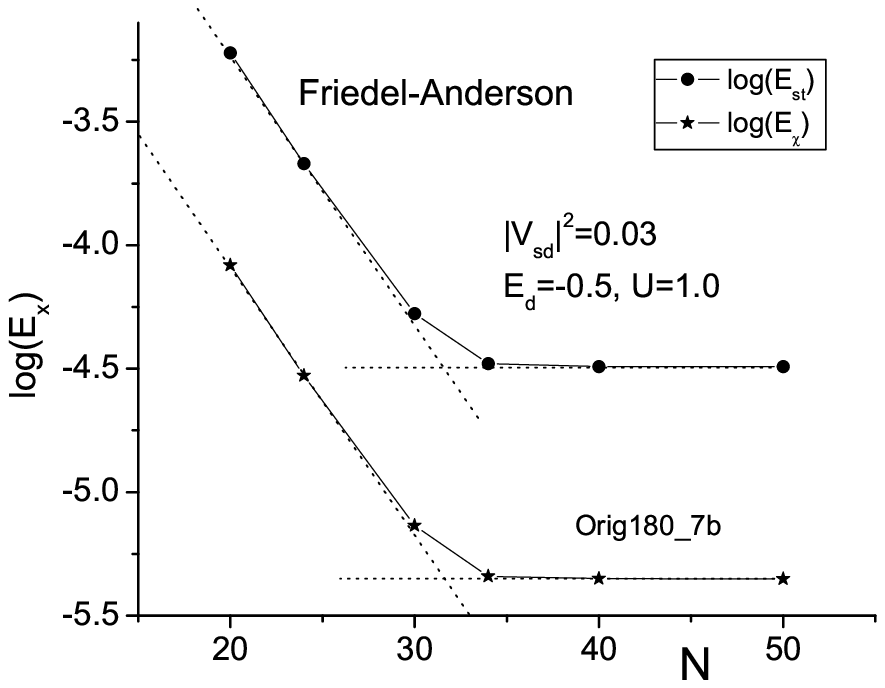}%
}%
%EndExpansion
\end{array}
\\
&
\begin{tabular}
[c]{l}%
Fig.12: The (logarithm of the) singlet-triplet excitation energy $E_{st}$\\
and the susceptibility energy $E_{\chi}=1/4\chi$ are plotted versus the\\
number $N$ of Wilson states. For $N$ larger than a a critical number\\
$N_{cr}\thickapprox33$ these energies are constant but below $N_{cr}$ both
energy\\
values increase.
\end{tabular}
\end{align*}%
\[
\]

\end{document}